**Global Cryptodemocracy is Possible and Desirable**
**By Ehud Shapiro (Weizmann Institute of Science)**

[A contribution to a Globalcit debate, April 5th, 2018]

The fascinating discussion kicked-off by Liav Orgad addresses the interplay between the clouds and earth: How do cloud citizens and cloud communities relate to their earthly counterparts?
Arguments by Orgad, Primavera De Filippi, Francesca Strumia, Peter Spiro and Dora Kostakopoulou espouse the potential benefits of global citizenship, ordained by the clouds, and cloud communities that such global citizens can form, inhabit and govern. Counterarguments by Rainer Bauböck, Robert Post, Michael Blake, Costica Dumbrava, Yussef Al Tamimi, Jelena Dzankic, Lea Ypi and Dimitry Kochenov suggest that what happens in the cloud stays in the cloud, and may not be helpful or relevant to, or at least cannot substitute for, earthly dominions, due to fundamental differences between the two. I will try to counter these counterarguments.
A key introductory point made by Bauböck is that Orgad "must have some form of global federal democracy in mind", yet that "his main vision is, however, the emergence of alternative forms of political community at the sub-global level". It is this main vision of Orgad that much of the weighty and thoughtful criticism is directed at.
To address it, I recall a strategy from mathematics: When faced with a difficult problem, namely a difficult theorem to prove, turn it into an even bigger problem: Define a more general and broader theorem, prove it, and then the original theorem easily follows as a corollary. This seemingly-paradoxical strategy works sometimes since a higher vantage point may offer a clearer view of the crux of the matter. I try to apply this strategy here: I will not address criticisms directed at sub-global political cloud communities directly. Instead, I will paint a vision of a global democracy, enabled by the Internet and the emerging technologies of blockchain and cryptocurrencies, explain how subsidiary communities based on shared territory or common interests, as envisioned by Orgad, can emerge and operate within it, and respond to criticism from this broader and more encompassing perspective.
From the outset, key criticisms that apply to subsidiary cloud communities do not apply to a global democracy, whether on or off the cloud (we note in parenthesis the respective critics): It has a clear territory (Bauböck, Post, Blake, Al Tamimi) – Planet Earth; it has diverse membership (Bauböck, Blake) – humanity at large; membership is involuntary (Bauböck, Post, Al Tamimi, Ypi) and by decree – just as earthly states conscript citizens by decree; it has room for political communities "that differ profoundly in their interests, identities and ideas about the common good" (Bauböck, Al Tamimi); and, due to all the above, it is clearly political (Bauböck, Post, Blake, Al Tamimi). Key remaining criticisms not answered by generalising the vision to incorporate all of humanity are those related to the use of coercion in community governance (Bauböck, Post, Blake, Dumbrava, Al Tamimi), lack of inclusivity (Ypi and Kochenov), and the risks of new technology (Dumbrava), which I will answer now in turn.
For our envisioned global democracy to be worthy of its name, it must uphold democratic values, including *sovereignty, equality, freedom of assembly, the*

*subsidiarity principle*, *transparency*, and the conservation of the natural and imprescriptible human rights: *liberty, property, safety and resistance against oppression*.

A fundamental advantage of blockchain technology is that it is the only technology to date that can uphold *sovereignty*: The multitudes participating in the operation of the blockchain are its sovereign; no member, third party or outside entity has omnipotent "super user" or "administrator" capabilities over the system, and no-one can pull the plug on it: it will survive as long as there are interconnected participants who are able and willing to continue its operation.[1] Hence, the answer to Stefania Milan's question, "do we really need the blockchain to enable the emergence of cloud communities?", is: Yes, if we want cloud communities to be sovereign and not subservient.

The situation is not as rosy with *equality*. Governance trepidations of the "cloud communities" of the leading cryptocurrencies, Bitcoin and Ethereum, which consist of their developers, miners and owners, resulted in community breakups termed "hard forks". Hence, second– and third-generation cryptocurrencies attempt to address their self-governance from first principles. However, they offer only plutocratic solutions, espousing "one coin – one vote" instead of the "one person – one vote" principle necessary for *equality*. It may be ironic, given the thrust of our discussion, that the only approach available today to realize equality on the blockchain is to piggyback on identities issued by earthly governments. Besides defeating the purpose of freeing cloud communities from the grasp of their earthly counterparts, this approach cannot mix and match governments or identity-granting authorities, lest people with multiple government-issued identities have multiple votes in the cloud; and it excludes people, such as refugees, who may be hard pressed to present a government-issued identity.

Realising truthful, unique and persistent global digital identities for all, a precondition for making an egalitarian blockchain, is a major open challenge.[2] But, for the sake of the vision we wish to paint, please suspend disbelief and assume that: (i) a worthy method for granting global digital identities to all has been devised, allowing any individual to claim a global identity (which functions as the "attested individual identities" Kochenov aspires for); call the rightful owners of such global identities *global citizens*; (ii) unhindered Internet access has been globally recognised as a basic civil right and is provided, directly or via a proxy, to any individual wishing to become a global citizen. While disbelief regarding the first assumption could be discharged in a decade, the second one will take longer. However, stating the goal of universal access as a basic civil right, taking concrete steps to implement it effectively, and making interim amends to compensate for its temporary lack, are all essential for our vision to be legitimate (and to address the justified criticisms of exclusion by Ypi and Kochenov). With this in mind, let us explore the vision of bringing about a global democracy of global citizens.

As much as disbelief is suspended, a method for granting global digital identities will never be perfect. Hence, the global democracy will have to grapple with fraud (fake, duplicate and stolen identities, Sybil attacks), extortion (the $5 wrench attack) and negligence (lost/forgotten password). Resolving such matters with due process would require a court. Such a court would need to rule according to a constitution. And the operation of the court (populated most likely by a combination of people

and machines) will have to be financed. So we have hardly left the doorstep in our journey towards a global cloud democracy, and already discovered that in order to realise *equality* we need a global court, a global constitution, and a global currency. That the global democracy needs a currency immediately suggests a cryptocurrency. But, how can we entrust the future of humanity to the hands of an environmentally-harmful[3], plutocratic regime? The answer is of fundamental importance: Current cryptocurrencies were architected on the premise that participants are anonymous and trustless, and resorted to the deliberately wasteful (Milan) proof-of-work protocol to cope with trustlessness. If indeed we have a mechanism for granting truthful and unique global digital identities that is reasonably resilient to attacks (e.g. at most one third of the global identities are compromised at any time) then the global democracy can deploy an egalitarian and planet-friendly cryptocurrency with a democratic governance regime; let's call such a cryptocurrency a *democratic cryptocurrency*.

Let's take stock: We have a democratic cryptocurrency governed by sovereign global citizens that are subject to a global court that rules according to a global constitution and is financed by the democratic cryptocurrency. This may sound a bit circular, but that's exactly how earthly states finance their operation. For example, the democratic cryptoeconomy can be fueled by a universal basic income to all global citizens. Income, wealth and transactions could be taxed, progressively if the global democracy decides so. Tax revenues would be disbursed to finance the operation of the global democracy, in particular the court and the underlying computational infrastructure ("mining"), as well as other purposes, according to a democratically-formed budget. To prevent speculative manipulation of the exchange rate of the democratic cryptocurrency, a global central bank may be established, with authority to purchase and sell foreign (crypto)currency to hinder such manipulations; the bank can similarly set an interest rate. The constitution will have to be updated as the global democracy develops, and subsidiary legislation will have to be adopted. So, in just a few short paragraphs we have come to realise that the global citizens of a global cloud democracy that has its own cryptocurrency and cryptoeconomy will have to recreate almost all the functions of earthly states; let's call this resulting specific vision a *global cryptodemocracy*, to distinguish it from the more general and abstract idea of a global democracy. If successful, it would show that a technology built with an "underlying philosophy of distributed consensus, open source, transparency and community" can be both "highly disruptive" *and* "serve similar purposes as those of states" (Milan); and it could achieve that without a reliance on the private sector and corporate capital that would necessitate paying undue attention to their interests and lobbying (Milan).

Additional key criticisms concern the ability of our global cryptodemocracy to protect human rights (Bauböck, Blake, Kochenov), collect taxes (Bauböck, Post) and in general enforce the rule of the law, given that physical coercion is possible on earth but not in the clouds (Bauböck, Post, Blake, Dumbrava, Al Tamimi). To redress crimes against global identities, we propose that global identities be realised as programmable software agents, aka "smart contracts", programmed to obey certified court orders. Thus, coercion is achieved through design and programmability, without violence: If the court determines a global identity to be fake, then it can directly order it to terminate; if determined to be a duplicate, then it can be ordered to merge into another identity, and if stolen then to change its owner. Regarding [Milan](#)'s observation that "activism today is characterised by […] a tendency to privilege flexible, multiple identities", we cannot hold the stick at both ends: aspire for egalitarian rule of law in a global democracy, and undermine it with flexible (and hence unaccountable) and multiple (and hence unfairly privileged) identities.

We propose to integrate the global citizen's global identity with her democratic [cryptocurrency wallet](#) into one entity, termed *global persona*. A global persona is the global citizen's proxy in the cloud: it is entrusted with the global citizen's identity information and crypto-assets, and it performs financial transactions and civic duties in the global cryptodemocracy on behalf of the global citizen it represents. Being unique and persistent makes a global persona accountable for the global citizen it represents. Hence, in addition to the court orders described above, a court may also issue fines against a global persona, payable immediately from her wallet, or deducted from her future (universal basic) income. As the global persona is programmed to obey court orders, no force is needed to collect such fines either. Income, wealth and transaction taxes can be similarly collected without the use of force, by programming global personas to obey the (democratically instituted) tax rules that are in effect. Of course, the court must be open to appeals on any decision and transaction.

A key remaining criticism relates to relying on and overseeing the technologies that will underlie our envisioned global cryptodemocracy ([Post](#), [Dumbrava](#)). The criticism is valid, but is mostly equally valid of any technology on which humanity depends today, and there are many. Perhaps one key technological vulnerability is related to the democratic process itself, ensuring that elections and more generally voting on the blockchain at least stand up to [earthly standards](#). Regarding overseeing blockchain technology, blockchain governance is indeed an issue of active research and experimentation, with the recognition that a change of underlying technology of a blockchain is as akin to, and as grave as, a change of constitution in a democracy. The global cryptodemocracy would employ the constitutional approach to its core technology, allowing constitutional change by its sovereign global citizens via a democratic process. Such a process must dampen the immediacy of Internet communication, lest mob dynamics may rule, by employing [hysteresis measures](#) such as special majority requirements. Recovery mechanisms would also be established, and invoked, by democratic decision.

Let us now consider [Orgad](#)'s vision of multiple Cloud Communities with a shared concern or ascriptive, thematic or geographic memberships "whose aim is political decision-making and in which individuals take part in a process of governance and the creation of law."

First, we note that all these communities can be subsidiary communities of the global cryptodemocracy, potentially with multiple levels of hierarchy (e.g. subsidiary animal rights or Bahá'í communities, with their own subsidiary communities based on country of residence); that the ability to form them is a manifestation of *freedom of assembly* in the clouds; and that allowing them to conduct their affairs without outside intervention is in line with the*subsidiarity principle*.

Second, such communities, within the context of a functioning global cryptodemocracy, may have at least one clear political goal: To draft and promote, within the parent global cryptodemocracy, policy and legislation that pertain to the rights and goals of their (possibly minority) community members. Recall the second article of the [1789 Declaration of the Rights of Man and Citizen](): "The goal of any political association is the conservation of the natural and imprescriptible rights of man. These rights are *liberty, property, safety and resistance against oppression*". To uphold these, the conduct of all subsidiary cloud communities must be *transparent* in order to ensure that no subsidiary community aims to harm the liberty, property or safety of other communities or global citizens.

Third, within these rich and multi-faceted cloud communities, a virtual punishment with a global scope against one's global persona, e.g. temporary suspension or even just a public reprimand, applied to all subsidiary cloud communities, would be severe indeed. Hence, the higher the value of the subsidiary cloud communities to peoples' lives, the mightier the coercive power of the global cryptodemocracy.

While we have implicitly assumed an egalitarian, democratic decision-making process at the core of global cryptodemocracy and in its subsidiary communities that will choose to adopt it, we have not specified this process. Such a mechanism faces many challenges, including "tyranny of structurelessness", "tyranny of emotions", decision-making by "microconsensus" within small cliques ([Milan]()) and many others. The question of how to best reach a democratic decision has been investigated sporadically for centuries (e.g. by [Llull](), [Condorcet](), [Borda]()), and intensively for the last 70 years within [Social Choice theory](). Much theory was developed, much confusion was sowed, and confidence in democracy has eroded, mainly due to [Arrow's impossibility theorem]() and its follow-on work. I will just hint that adding [a taste for reality to social choice]() theory can undo much of this damage and restore trust in democratic decision making, on and off the cloud.

I have aimed to show that a vision of a global cryptodemocracy, with a rich set of subsidiary cloud communities, is realisable and have tried to address many of the criticisms raised in this debate. But, even if a global cryptodemocracy is realisable, and successfully addresses criticism, is it desirable? My personal answer is positive for two reasons: First, I believe that, [since the days of Kant and even before](), the proponents of a world government own the moral high ground, and the weakness of their position was practical: Until now, for a world government to materialise, local governments have to volunteer to cease some of their power; and giving up of power is not known to happen voluntarily. Fortunately, earthly democracies are sufficiently free so that the formation of a global cryptodemocracy does not require their consent. True, dictatorial regimes may prevent their citizens from participating, but this would, eventually, be at their own peril, as the interests of their people will not be represented as well. And true, the full power of a global cryptodemocracy will not be realised until proponents of global democracy become majorities in the majority of their respective earthly states. Yet, embryonic as it may be, the global cryptodemocracy vision presented here may very well be the only concrete proposal towards the ultimate realisation of a global democratic government based on currently available technologies.

And this relates to my second reason. I believe that for representative democracies to rebounce from their worldwide decline, they should undergo a major revision and adopt the practices of one of the oldest and most successful democracies in the world, namely the Swiss federal direct democracy. Given that those in power never give it up voluntarily, and that direct democracy disempowers representatives, such a major shift cannot happen without a major outside force in its favor. And new technology can offer such a force. In particular, political e-parties, formed as subsidiary cloud communities of the global cryptodemocracy, sharing the same technology and networking to share winning practices and methods, may be able to win earthly elections and change earthly democracies for the better. This in turn may result in such earthly democracies officially supporting[4] the global cryptodemocracy in its rise into a *bona fide* egalitarian democratic world government of all global citizens.

**Notes:**
[1] I acknowledge Milan's point that such interconnectedness (but not the computers being connected!) would most-probably be commercially-owned, and that it is essential that such interconnectedness be neutral and unhindered, even if owned and controlled by private or government interests. Given that, global citizens can be the true sovereign of the global democratic blockchain outlined below.
[2] Disclosure: My team at Weizmann aims to address this global challenge. Note that it will not be solved just by achieving broader coverage of local government-issued IDs.
[3] Present-day cryptocurrencies are unsustainable, even environmentally-harmful, since the proof-of-work protocols that underlie, for example, Bitcoin and Ethereum are unfathomably energy-wasteful on purpose: The ongoing operation of Bitcoin alone consumes as of today more energy than does the entire state of Israel, with its more than 8 million inhabitants.
[4] For example, a state may create government-attested global personas for all its citizens, place them in the escrow of the state notary, and assign them to citizens upon their presentation of a government-issued ID. This would immediately turn all state citizens into global citizens. A state citizen who already owns a global persona will have to merge it with the received government-attested global persona, lest she would be guilty of owning duplicate global personas.